\documentclass{article}

\usepackage{arxiv}

\usepackage[utf8]{inputenc} % allow utf-8 input
\usepackage[T1]{fontenc}    % use 8-bit T1 fonts
\usepackage{hyperref}       % hyperlinks
\usepackage{url}            % simple URL typesetting
\usepackage{booktabs}       % professional-quality tables
\usepackage{amsfonts}       % blackboard math symbols
\usepackage{nicefrac}       % compact symbols for 1/2, etc.
\usepackage{microtype}      % microtypography
\usepackage{lipsum}		% Can be removed after putting your text content
\usepackage{graphicx}
\usepackage{natbib}
\usepackage{doi}
\usepackage{array}
\usepackage{longtable}

\title{FairFlow Protocol: Equitable Maximal Extractable Value (MEV) mitigation in Ethereum}

\author{ \href{https://orcid.org/0000-0001-5431-6367}{\includegraphics[scale=0.06]{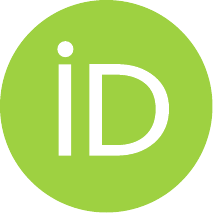}\hspace{1mm}Dipankar Sarkar} \\
  Cryptuon Research \\
  \texttt{me@dipankar.name} \\
}

\hypersetup{
pdftitle={FairFlow Protocol: Equitable Maximal Extractable Value (MEV) mitigation in Ethereum},
pdfsubject={cs.CR},
pdfauthor={Dipankar Sarkar},
pdfkeywords={Ethereum, MEV, Mitigation, Game Theory},
}

\begin{document}
\maketitle

\begin{abstract}
Ethereum has become a popular platform for decentralized applications (dApps) due to its powerful smart contract capabilities. Nevertheless, Maximal Extractable Value (MEV) has been a major issue in the Ethereum ecosystem, with considerable implications for the platform's functioning and integrity. This paper introduces the FairFlow protocol, a new system created to address the MEV problem within Ethereum's existing infrastructure. The protocol seeks to create a fairer environment, preventing miners or validators from taking advantage of the system, and safeguarding user data. The combination of auction-based block space allocation and randomized transaction ordering significantly reduces the possibility of MEV exploitation.
\end{abstract}

% keywords can be removed

\section{Introduction}
The advent of blockchain technology with Blockhain\citep{nakamoto2008bitcoin}, epitomized by platforms such as Ethereum \citep{wood2014ethereum}, has significantly transformed digital transactions. This transformation is mainly attributed to the introduction of key features such as decentralization, transparency, and immutability. Ethereum has been particularly prominent as a platform for decentralized applications (dApps) due to its pioneering smart contract functionality \citep{buterin2014ethereum}.

However, Ethereum's operational mechanics, including transaction processing \citep{etherscan2023}  and block validation, present specific challenges \citep{ethereumfoundation2020}. A notable issue within the Ethereum ecosystem is Maximal Extractable Value (MEV) \citep{daian2019flash}. MEV refers to the maximum profit that miners or validators can make by ordering transactions within a block, which is above standard block rewards and gas fees. This phenomenon has led to practices such as front-running, back-running, and transaction reordering, which could undermine fairness and efficiency in the network \citep{qin2020quantifying}.

The implications of MEV are vast, affecting aspects such as transaction prioritization, network congestion, and the overall trustworthiness of the Ethereum platform \citep{daian2019flash} \citep{qin2020quantifying}. To mitigate the issue of MEV, several solutions have been proposed. These include the implementation of off-chain transaction ordering services \citep{buterin2021rollup}, alterations in consensus protocols, and the introduction of more robust transparency mechanisms \citep{eipgithub}. Although promising, these solutions have their limitations, often introducing new complexities or failing to address the core problem in a comprehensive way \citep{gudgeon2020sok}. For example, off-chain services might inadvertently lead to centralization in transaction ordering, and changes to consensus protocols could require significant alterations to the Ethereum framework, which pose substantial implementation challenges \citep{ethereumfoundation2020}.

This paper introduces the FairFlow protocol, a framework created to tackle the issues of Maximal Extractable Value (MEV) in the Ethereum environment. As an additional layer to Ethereum, FairFlow provides a set of mechanisms that prioritize fairness and privacy in transaction processing. The protocol utilizes advanced cryptographic techniques and a distinct decentralized auction system for transaction ordering.

This innovative approach not only curtails the potential for MEV exploitation but also fortifies the foundational principles of Ethereum, including decentralization and openness, with a renewed emphasis on equitable outcomes for all network participants.The contributions of this paper to the discourse on Ethereum's MEV challenges are manifold and substantial.

\begin{itemize}
    \item FairFlow is a new layered architectural model that has been specifically designed to improve fairness in transaction processing without changing the core architecture of Ethereum.
    \item Exploring the intricate components of FairFlow, demonstrating its smooth connection to the Ethereum network, and emphasizing how it safeguards and reinforces Ethereum's commitment to fairness.
    \item This paper examines the impact of FairFlow on transactional equity, privacy, and network efficiency, with the aim of preventing any single entity from monopolizing the advantages.
    \item Gain important understanding of the lasting effects of executing equitable MEV mitigation approaches, thereby forming the course of forthcoming progressions in decentralized blockchain settings.
\end{itemize}

This paper puts FairFlow at the cutting edge of attempts to adjust the Ethereum network, shifting towards a blockchain environment that is more equitable and inclusive.

\section{Literature review}

Since its inception, Ethereum has been a pioneering force in blockchain technology. The Ethereum whitepaper, published in 2014 by Vitalik Buterin, laid the groundwork for the platform, which was officially launched in 2015 \citep{buterin2014ethereum}. Ethereum has evolved significantly over the years, marked by its community-driven open-source development. In particular, the transition from a Proof of Work (PoW) to a Proof of Stake (PoS) consensus mechanism marks a pivotal moment for the platform, which affects its scalability, security and energy efficiency \citep{buterin2020ethereum2}.

The literature on Ethereum, MEV, blockchain privacy, and fairness is extensive, yet evolving. It underscores the challenges and opportunities for innovation within Ethereum and similar platforms. Despite significant research, there remain gaps in effectively mitigating MEV, improving transaction privacy, and ensuring fairness. FairFlow aims to address these gaps, offering a novel approach that integrates enhanced transaction privacy and fairness into Ethereum's existing infrastructure.

\subsection{Maximal Extractable Value (MEV)}

Maximal Extractable Value (MEV) has emerged as a critical concept in the Ethereum ecosystem. MEV is the maximum profit a miner or validator can extract from manipulating the order of transactions within a block, in addition to the rewards of the block and the transaction fees \citep{daian2019flash}. As the Ethereum ecosystem has expanded, MEV has evolved from a theoretical concern to a practical challenge with significant implications for platform integrity and operation.

MEV impacts Ethereum's network performance and user experience, leading to network congestion and fluctuating transaction fees \citep{qin2020quantifying}. Practices associated with MEV, such as front-running, pose security and fairness issues, potentially leading to financial losses and a loss of trust in the platform \citep{eskandari2019bitcoinkey}.

To mitigate MEV, solutions such as Flashbots have been introduced, offering a private channel for users to negotiate transaction ordering with miners \citep{flashbots2021mev}. Furthermore, chain adjustments, such as EIP 1559, aim to make transaction fees more predictable \citep{buterin2021rollup}.

\subsection{Transaction Privacy}

The privacy of blockchain transactions remains a fundamental concern. Despite the inherent transparency of blockchains, the need for privacy, especially in financial transactions, is crucial. Techniques for improving privacy include address mixing and advanced cryptographic methods such as Zero-Knowledge Proofs (ZKP) \citep{bensasson2014zerocash}. ZKPs enable transaction validation without disclosing transaction details, enhancing privacy.

\subsection{Transaction Ordering}

Fair transaction ordering is critical to blockchain fairness. In an ideal system, transactions would be processed in the order received. However, miners often have discretion over the transaction order, which can lead to unfair outcomes \citep{bonneau2015bitcoinresearch}. Proposals for fair transaction processing range from randomized ordering to consensus algorithms that consider transaction fairness \citep{kiayias2017ouroboros}.

\section{FairFlow Protocol Design}

FairFlow is designed to address critical challenges in the Ethereum ecosystem, focusing on mitigating miner extractable value (MEV), improving transaction privacy, and ensuring fairness in transaction processing. The protocol aims to provide a \textit{more equitable environment}, preventing exploitation by miners or validators, and protecting user data.

\paragraph{Integration with Ethereum} The processed transactions are seamlessly integrated into Ethereum blocks. FairFlow works in tandem with Ethereum's consensus mechanism, maintaining compatibility and interoperability.

FairFlow's architecture inherently reduces MEV opportunities. The combined approach of auction-based block space allocation and randomized transaction ordering significantly reduces the potential for MEV exploitation.

The protocol enhances the privacy of transactions beyond Ethereum's current capabilities. By encrypting transactions and carefully managing their exposure, FairFlow offers a higher level of privacy.

\begin{figure}[h]
    \centering
    \includegraphics[width=0.8\linewidth]{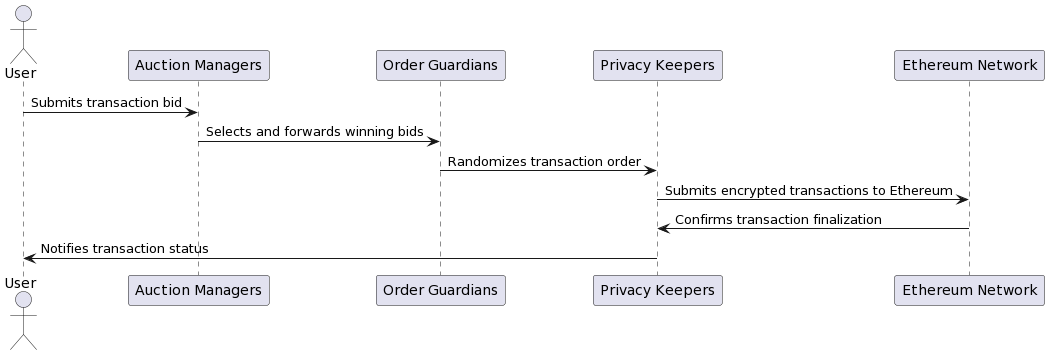}
    \caption{High-level Flow diagram}
    \label{fig:enter-label}
\end{figure}

\subsection{Specialised Nodes}

FairFlow introduces an extra layer to the existing Ethereum architecture, which acts as a sophisticated intermediary. This layer captures, processes, and then sends transactions to the Ethereum network, providing improved security, privacy, and fairness. FairFlow's network structure consists of various specialized nodes, each of which has an essential role.

\begin{itemize}
    \item \textbf{Auction Managers} These nodes manage the auction process for block space allocation. Their primary function is to ensure that the allocation of block space is fair and transparent.
    \item \textbf{Order Guardians} They are responsible for randomly deciding the transaction order, thus mitigating the potential exploitation of MEVs.
    \item \textbf{Privacy Keepers} Employing advanced cryptographic techniques, these nodes ensure the confidentiality of transaction details until they are irrevocably recorded on the blockchain.
\end{itemize}

\subsection{Block Auction Mechanism}

\begin{figure}[h]
    \centering
    \includegraphics[width=0.4\linewidth]{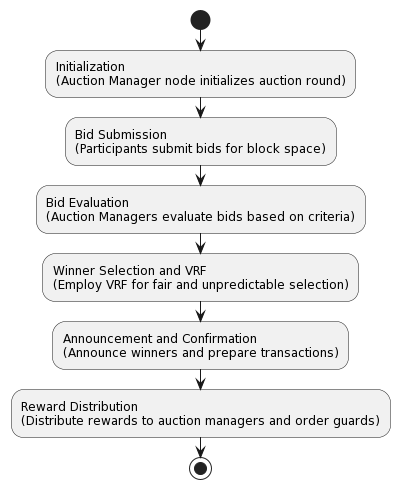}
    \caption{Auction flow}
    \label{fig:enter-label}
\end{figure}

FairFlow utilizes an auction mechanism as a key element. Auction managers are responsible for running a bidding process for block space, deciding on the winners based on predetermined criteria. This process includes submitting bids, deciding on the winners, and distributing rewards.

The auction system is designed to guarantee fairness in the inclusion of transactions, eliminating the possibility of preferential treatment and creating an even playing field for all participants.

\begin{enumerate}
    \item \textbf{Initialization} Each Auction Manager node initializes an auction round for block space allocation. This round is synchronized with Ethereum's block production schedule to ensure seamless integration.
    \item \textbf{Bid Submission} Participants (i.e., transaction senders or their representatives) submit bids for block space. These bids can include factors like the transaction fee offered, the urgency of the transaction, and other relevant metadata.
    \item \textbf{Bid Evaluation} Auction Managers evaluate these bids based on predefined criteria. These criteria are transparent and consistent across the network, ensuring fairness.
    \item \textbf{Winner Selection and VRF} To enhance fairness and unpredictability in winner selection, FairFlow employs a Verifiable Random Function (VRF). VRF provides cryptographic proof that the selection process is random and tamper-proof, preventing any form of manipulation.
    \item \textbf{Announcement and Confirmation} Once winners are selected, the Auction Managers announce the results. Winning transactions are then prepared for inclusion in the next Ethereum block.
    \item \textbf{Reward Distribution} After successful block creation, the rewards (transaction fees) are distributed to auction managers and order guards, adhering to the protocol’s reward distribution mechanism.
\end{enumerate}

\subsection{Transaction Handling}

\begin{figure}[h]
    \centering
    \includegraphics[width=0.4\linewidth]{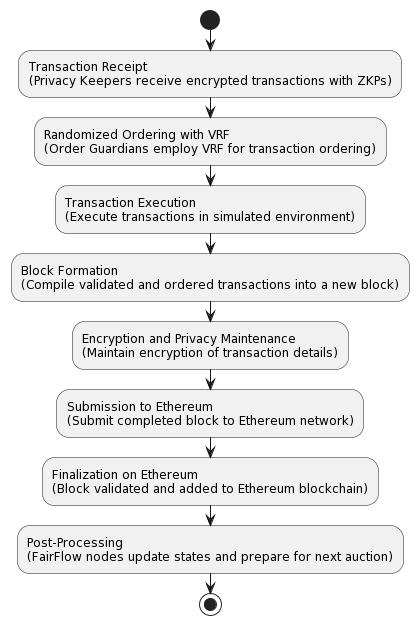}
    \caption{Transaction flow}
    \label{fig:enter-label}
\end{figure}

FairFlow transactions are secured through the use of advanced cryptographic techniques, including zero-knowledge proofs (ZKPs). This encryption guarantees that the details of the transaction remain confidential until they are executed on the blockchain.

Order Guardians in FairFlow randomize the order of transactions. This randomness is crucial to mitigate MEV risks, to ensure that no predictable patterns can be exploited.

\begin{enumerate}
    \item \textbf{Transaction Receipt} Privacy Keepers receive encrypted transactions from users. These transactions are encrypted using advanced cryptographic techniques, including Zero-Knowledge Proofs (ZKPs).
    \item \textbf{Randomized Ordering with VRF} Order Guardians receive the set of transactions selected by the Auction Managers. They then employ VRF to determine the random order of these transactions. This process ensures that the transaction order within a block is unpredictable, thus mitigating MEV risks.
    \item \textbf{Transaction Execution} The randomly ordered transactions are executed in a simulated environment to predict their outcomes. This step is crucial for ensuring that only valid transactions are included in the block.
    \item \textbf{Block Formation} Once transactions are validated and ordered, they are compiled into a new block. This block adheres to Ethereum’s block structure but with the added layer of FairFlow’s processing.
    \item \textbf{Encryption and Privacy Maintenance} During the block formation process, transaction details remain encrypted. Privacy Keepers ensure that only minimal necessary information is exposed, maintaining user privacy until the final execution on the Ethereum blockchain.
    \item \textbf{Submission to Ethereum} The completed block is then submitted to the Ethereum network. This submission is timed to align with Ethereum's block production cycle, ensuring a smooth integration.
    \item \textbf{Finalization on Ethereum} Upon successful validation by Ethereum's consensus mechanism, the block is added to the Ethereum blockchain. This finalization marks the completion of the transaction and auction process within FairFlow.
    \item \textbf{Post-Processing} After block finalization, FairFlow nodes engage in post-processing activities. This includes updating internal states, preparing for the next auction round, and executing any necessary protocol-level maintenance tasks.
\end{enumerate}

\section{Protocol Analysis}

The introduction of a system like FairFlow alongside Ethereum, especially one that encompasses encrypted transactions and additional measures to mitigate Maximal Extractable Value (MEV), presents several benefits over Ethereum's regular operation. 

\subsection{System Benefits}

This system is potentially advantageous for different stakeholders in the Ethereum ecosystem, including users, validators, developers, and the broader community. We outline the primary benefits and incentives for these participants in table \ref{table:fairflow}.

\begin{table}[h]
\centering
\begin{tabular}{ | m{6cm} | m{6cm} | }
\hline
\textbf{Users} & \textbf{Validators/Miners} \\
\hline
\textbf{Enhanced Privacy}: Encrypted transactions within blocks ensure additional privacy. & \textbf{New Revenue Streams}: Possible novel avenues for validators to earn rewards. \\
\textbf{Reduced MEV Risks}: Auction-based block space allocation and random transaction ordering reduce front-running risks. & \textbf{Reduced Ethical Dilemmas}: Focus on network security rather than exploiting transaction sequencing. \\
\textbf{Fairer Transaction Processing}: Transparent process for transaction inclusion. & \textbf{Increased Network Trust and Value}: Greater trust and potentially higher ETH valuation. \\
\hline
\textbf{Developers} & \textbf{Ethereum Community} \\
\hline
\textbf{Broader Design Space}: Opportunity to create dApps utilizing FairFlow's unique features. & \textbf{Enhanced Network Integrity}: Tackling MEV issues and advancing transaction privacy. \\
\textbf{Innovative Opportunities}: Challenges in dApp development and smart contract design. & \textbf{Attracting Diverse Users}: Appealing to a broader spectrum of users. \\
& \textbf{Setting Industry Standards}: Establishing new benchmarks in blockchain technology. \\
\hline
\end{tabular}
\caption{FairFlow Benefit Analysis}
\label{table:fairflow}
\end{table}

Integration of a system like FairFlow into the Ethereum network brings multidimensional advantages, aligning with the diverse interests of stakeholders within the Ethereum ecosystem. Addresses some of Ethereum's inherent challenges, especially concerning MEV and transaction privacy, thus enhancing the network's overall value proposition. 

This integration could make Ethereum an increasingly attractive and robust platform for a wide range of decentralized applications, financial services, and other blockchain-based utilities. 

However, implementing and adopting such a system requires careful coordination, consensus within the community, and a balanced approach to incentives for all parties involved.

\subsection{Game Theoritic Analysis}

In order to assess the Nash Equilibrium for Auction Managers, Order Guardians, and Privacy Keepers in the FairFlow architecture, we must take into account their strategic decisions and the rewards that come with them. We will assume that all participants are rational and are looking to maximize their own benefits.

\begin{table}[h]
\centering
\begin{tabular}{ | m{2cm} | m{6cm} | m{6cm} | }
\hline
\textbf{Role} & \textbf{Strategy and Payoff} & \textbf{Nash Equilibrium} \\
\hline

\textbf{Auction Managers} & 
\textbf{Strategy:} Choose to run fair auctions or collude. \textbf{Payoff:} Fair auctions ensure steady income and reputation. Collusion offers immediate gain but risks detection and loss of position. & 
In equilibrium, they choose to run fair auctions due to the risk of losing their position outweighing potential immediate gains. \\
\hline

\textbf{Order Guardians} & 
\textbf{Strategy:} Randomize transaction order genuinely or manipulate for benefits. \textbf{Payoff:} Genuine randomization maintains role and reputation. Manipulation offers immediate benefits but carries high detection risk. & 
The equilibrium strategy is genuine randomization, as the risk of detection and penalties are too high compared to short-term gains. \\
\hline

\textbf{Privacy Keepers} & 
\textbf{Strategy}: Maintain encryption reliably or leak information for benefit.\textbf{ Payoff:} Reliable maintenance ensures job security and income. Leaking poses high risk of detection and loss of future earnings. & 
They will maintain encryption reliably, as the risk of unethical behavior is unattractive compared to their reputation and consequences. \\
\hline
\end{tabular}
\caption{Analysis of Roles in FairFlow Architecture}
\label{table:fairflow-roles}
\end{table}

The Nash Equilibrium is in line with ethical and responsible conduct due to the high risk of unethical behavior, combined with the transparency of the system and efficient monitoring systems. The FairFlow architecture, which includes checks, balances, and transparency, encourages participants to act in the best interest of the network, making individual incentives compatible with the overall health and integrity of the network. This equilibrium is only possible if the system is able to detect and punish misconduct, which is essential for maintaining these equilibrium strategies.

\subsection{Comparative Analysis}

The Ethereum blockchain landscape is constantly changing, and FairFlow is a major step forward in tackling the issues related to Miner Extractable Value (MEV). In comparison to MEV-Boost, which works to decentralize block production and distribution to reduce the negative effects of MEV, FairFlow offers a new solution. This additional layer to Ethereum increases transaction privacy and fairness, and directly addresses MEV worries. This distinction in purpose reflects the growing importance of privacy and fairness in the blockchain world. Table \ref{table:mev-fairflow} provides a comparison of the two protocols.

The differences between MEV-Boost and FairFlow are noteworthy. MEV-Boost divides block building from proposal, allowing validators to obtain blocks from multiple builders, thus reducing the risk of censorship and collusion. On the other hand, FairFlow introduces new roles such as auction managers, order guards, and privacy keepers, as well as encrypted transactions and randomized order execution. These features work together to reduce the exploitation of MEV by hiding transaction details until execution, thereby improving the network's integrity. MEV-Boost seeks to democratize MEV access and maintain network security, while FairFlow adds complexity but significantly increases transaction privacy and fairness. This shift in approach reflects a deeper understanding of the MEV phenomenon and a commitment to address its more subtle and insidious effects.

FairFlow's integration, though more intricate, provides a hopeful outlook for Ethereum's network in the future, unifying individual motivations with the general well-being and reliability of the network.

\begin{table}[h]
\centering
\begin{tabular}{ | >{\raggedright}p{3cm} | >{\raggedright}p{6cm} | >{\raggedright\arraybackslash}p{6cm} | }
\hline
\textbf{Criteria} & \textbf{MEV-Boost} & \textbf{FairFlow} \\
\hline

\textbf{Purpose} & Aims to mitigate the negative impacts of MEV in a post-merge Ethereum environment, focusing on decentralizing block production and distribution. & Provides an additional layer to Ethereum enhancing transaction privacy and fairness, addressing MEV concerns. \\
\hline

\textbf{Functionality} & 
Separates block building from the proposal. Validators receive blocks from multiple builders, reducing the risk of censorship. & 
Introduces roles like Auction Managers, Order Guardians, and Privacy Keepers. Implements encrypted transactions and randomized order execution. \\
\hline

\textbf{MEV Handling} & 
Allows fair access to MEV opportunities by choosing from various block builders. Validators select profitable MEV-optimized blocks. & 
Reduces MEV exploitation by encrypting transactions and randomizing their order. Conceals transaction details until execution. \\
\hline

\textbf{Impact on Network} & 
Democratizes MEV access, maintaining network security, and decentralization by reducing the collusion potential of the validator. & 
Adds complexity but enhances privacy and fairness in transactions, altering MEV dynamics on Ethereum. \\
\hline

\textbf{Approach to MEV} & 
Decentralizes block-building process, democratizing MEV opportunities. & 
Reduces MEV opportunity by encrypting transactions and randomizing order, focusing on privacy and fairness. \\
\hline

\textbf{Network Integration} & 
Direct plug-in solution for Ethereum's PoS system. & 
Comprehensive integration, adding new layers and roles to Ethereum. \\
\hline

\textbf{Complexity and Usability} & 
Maintains operational simplicity within Ethereum's framework. & 
Introduces new complexities and roles, altering Ethereum's fundamental functions. \\
\hline

\textbf{Privacy and Security} & Focus on equitable distribution of MEV and maintaining network decentralization and security. & 
Emphasizes transaction privacy and security against MEV. \\
\hline
\end{tabular}
\caption{Comparative Analysis of MEV-Boost and FairFlow}
\label{table:mev-fairflow}
\end{table}

\section{Challenges and Limitations}

FairFlow, while an innovative approach to improving the Ethereum network, encounters several challenges and limitations that merit attention. One significant concern is the added complexity; the incorporation of new layers and roles into the Ethereum framework could potentially complicate the user experience and hinder adoption rates. This complexity requires a balance between innovation and usability.

Moreover, the integration and compatibility of FairFlow with Ethereum's existing infrastructure pose substantial challenges. It requires meticulous design to ensure seamless functionality, especially considering Ethereum's ongoing and future upgrades. This aspect is critical for maintaining the system's integrity and avoiding disruptions in service.

Another vital consideration is scalability. Blockchain solutions, by their nature, face scalability challenges, and FairFlow is no exception. It is imperative that the introduction of FairFlow's functionalities does not impede the network's performance or scalability. This balance is crucial to ensure that the benefits of FairFlow do not come at the cost of reduced network efficiency.

Lastly, the introduction of new components and mechanisms within FairFlow brings forth new security considerations. Each addition to the system potentially opens up vulnerabilities that could be exploited. Therefore, ensuring the robustness and security of FairFlow against potential attacks is of utmost importance. This requires ongoing vigilance and adaptation to evolving security threats to safeguard the integrity and reliability of the network.

\section{Future Work}

As FairFlow development progresses, several key areas have emerged as critical for further exploration and advancement. A primary area of focus is the enhancement of cryptographic techniques. Continued research in this domain, particularly methods that improve privacy and security in decentralized networks, is vital for the fortification of FairFlow. Advanced cryptographic methods have the potential to significantly strengthen the security and integrity of the system.

Another crucial area of research is the scalability of FairFlow. Given the challenges associated with blockchain scalability, investigating and implementing solutions that could enhance the scalability of FairFlow is imperative. This exploration might include the adaptation of layer 2 solutions or sharding mechanisms, which are essential to ensure the long-term viability and success of the system.

Furthermore, a more in-depth economic analysis is necessary to fully understand the implications of FairFlow’s auction mechanism on transaction fees and validator incentives within the Ethereum network. Such an analysis would provide valuable information on the economic impact and efficiency of FairFlow, guiding future improvements and adaptations.

Exploring the potential of FairFlow in cross-chain applications also presents an exciting opportunity. Adapting FairFlow to function across different blockchain platforms could significantly expand its impact and utility within the decentralized ecosystem. This exploration would contribute to the broader application and influence of FairFlow in various blockchain contexts.

Lastly, a user-centric design approach is paramount. With the added complexity introduced by FairFlow, it is essential to ensure that the user experience remains intuitive and accessible. Future work should prioritize user-friendliness and accessibility, making FairFlow a practical and convenient solution for users within the Ethereum network and potentially beyond. This focus on user-centric design is crucial for achieving widespread adoption and usability of FairFlow.

\section{Conclusion}

This paper introduced FairFlow, a novel protocol designed to enhance transaction privacy and fairness while mitigating Maximal Extractable Value (MEV) in the Ethereum ecosystem. FairFlow's unique approach, integrating Auction Managers, Order Guardians, and Privacy Keepers into Ethereum's transaction processing, marks a significant step forward in addressing some of the most pressing challenges in blockchain technology. By implementing a transparent auction mechanism and employing advanced cryptographic methods, FairFlow aims to democratize transaction processing, reduce MEV opportunities, and protect user privacy.

FairFlow represents a meaningful contribution to ongoing efforts to make blockchain technology more secure, private, and fair. While challenges remain, the potential benefits of the protocol to the Ethereum community and the broader blockchain landscape are significant. As the field continues to evolve, FairFlow stands as a testament to the innovative spirit driving the blockchain community forward, continually seeking solutions that improve the efficacy, security, and fairness of decentralized networks.

\bibliographystyle{unsrtnat}
\bibliography{references}

\end{document}